\documentclass[prl,showpacs,twocolumn]{revtex4}

\begin{document}

\title{Concurrence of Arbitrary Dimensional Bipartite Quantum States}
\author{Kai Chen$^{1}$}
\author{Sergio Albeverio$^{1}$}
\author{Shao-Ming Fei$^{1,2}$}
\affiliation{$^1$Institut f\"ur Angewandte Mathematik, Universit\"at Bonn, D-53115,
Germany\\
$^2$Department of Mathematics, Capital Normal University, Beijing 100037,
China}

\begin{abstract}
We derive an analytical lower bound for the concurrence of a bipartite
quantum state in arbitrary dimension. A functional relation is established
relating concurrence, the Peres-Horodecki criterion and the realignment
criterion. We demonstrate that our bound is exact for some mixed quantum
states. The significance of our method is illustrated by giving a
quantitative evaluation of entanglement for many bound entangled states,
some of which fail to be identified by the usual concurrence estimation
method.
\end{abstract}


\maketitle

Entanglement is a striking feature of quantum systems and is the key
physical resource to realize quantum information tasks such as quantum
cryptography, quantum teleportation and quantum computation \cite{nielsen},
which cannot be accounted for by classical physics. This has provided a
strong motivation for the study of detection and quantification of
entanglement in an operational way. Despite of a great deal of effort in
past years \cite%
{Peres96,HorodeckiPLA96,Wootters98,Audenaert01,Rudolph02,ChenQIC03,Horodecki02,realignmentcriteria,chenp02-Gerjuoy03,Lozinski03,sepcriteria,Mintert04-MintertPhD}
for the moment only partial solutions are known for generic mixed states. As
for quantitative measures of entanglement, there is an elegant formula for 2
qubits in terms of \emph{concurrence}, which is derived analytically by
Wootters in Ref. \cite{Wootters98}. This quantity has recently been shown to
play an essential role in describing quantum phase transition in various
interacting quantum many-body systems \cite{Osterloh02-Wu04} and may affect
macroscopic properties of solids significantly \cite{Ghosh2003}.
Furthermore, value of concurrence will provide an estimation \cite%
{Mintert04-MintertPhD} for the entanglement of formation (EOF) \cite{BDSW},
which quantifies the required minimally physical resources to prepare a
quantum state. It is thus very important to have a precise quantitative
picture of entanglement in order to get a better insight into the
corresponding physical systems.

However, calculation of the concurrence is a formidable task as the Hilbert
space dimension is increasing, like in the case of two parts in a real
solid-state system considered for quantum computation. Good algorithms and
progresses have been obtained concerning lower bounds for qubit-qudit system
\cite{chenp02-Gerjuoy03,Lozinski03} and for bipartite systems in arbitrary
dimension \cite{Audenaert01,Mintert04-MintertPhD}. Considerable progress is made
in \cite{Mintert04-MintertPhD} to give a purely algebraic lower bound.
Nevertheless, an optimized bound generally involves numerical optimization
over a large number of free parameters in a level (at least $m(m-1)n(n-1)/4$
for a $m\otimes n$ bipartite system, where $m,n$ are Hilbert space dimension
for two subsystems respectively \cite%
{Audenaert01,Lozinski03,Mintert04-MintertPhD}). This leads to a
computationally untractable problem for realistic system with a higher
dimension. In addition, these methods for evaluating concurrence can not
detect reliably \emph{arbitrary} entangled states even if one applies all known
optimization methods \cite{Mintert04-MintertPhD}.

Our aim in this work is to improve this situation dramatically by giving an
analytical lower bound for concurrence of any mixed bipartite quantum state.
We find an essential quantitative relation among this measure and available
strong separability criteria. A functional relation is explicitly derived to
give a tightly lower bound for the concurrence. It is shown to be exact for
some special class of states. Our method is further demonstrated to be
better than the regular method for concurrence optimization, in the sense
that it can detect and give an evaluation of entanglement for many bound
entangled states (BES) which cannot be identified by the latter. This also
complements a number of existing methods involving numerical optimization
and provides a computational method to estimate manifestly the actual value
of concurrence for any bipartite quantum state.

We start with a generalized definition \cite{Rungta01-AlbeverioFei01} of
concurrence for a pure state ${\left\vert \psi \right\rangle }$ in the
tensor product $\mathcal{H}_A \otimes \mathcal{H}_B$ of two (finite
dimensional) Hilbert spaces $\mathcal{H}_A,\mathcal{H}_B$ for 2 systems $A,B$%
. The concurrence is defined by $C({\left\vert \psi \right\rangle })=\sqrt{%
2(1-\mbox{Tr}\rho _{A}^{2})}$, where the reduced density matrix $\rho _{A}$
is obtained by tracing over the subsystem $B$.
The concurrence is then extended to mixed states $\rho$ by the convex roof,%
\begin{equation}
C(\rho )\equiv \min_{\{p_{i},|\psi _{i}\rangle \}}\sum_{i}p_{i}C({\left\vert
\psi _{i}\right\rangle }),  \label{concurrence}
\end{equation}%
for all possible ensemble realizations $\rho =\sum_{i}p_{i}|\psi _{i}\rangle
\langle \psi _{i}|$, where $p_{i}\geq 0$ and $\sum_{i}p_{i}=1$. For any pure
product state ${\left\vert \psi \right\rangle }$, $C({\left\vert \psi
\right\rangle })$ vanishes according to the definition. Consequently, a
state $\rho $ is \emph{separable} if and only if $C(\rho )=0$ and hence can
be represented as a convex combination of product states as $\rho
=\sum_{i}p_{i}\rho _{i}^{A}\otimes \rho _{i}^{B}$ where $\rho _{i}^{A}$ and $%
\rho _{i}^{B}$ are pure state density matrices of the subsystems $A$ and $B$%
, respectively \cite{werner89}.

The key point of our idea is to relate directly the concurrence and the
Peres-Horodecki criterion of positivity under partial transpose (PPT
criterion) \cite{Peres96,HorodeckiPLA96} and the realignment criterion \cite%
{Rudolph02,ChenQIC03} by means of Schmidt coefficients of a pure state. Let
us firstly consider the concurrence for a pure state. $C({\left\vert \psi
\right\rangle })$ is invariant under a local unitary transformation (LU)
\cite{Wootters98,Rungta01-AlbeverioFei01}. Without loss of generality, we
suppose that a pure $m\otimes n$ $(m\leq n)$ quantum state has the standard
Schmidt form
\begin{equation}
{\left\vert \psi \right\rangle }=\sum_{i}\sqrt{\mu _{i}}{\left\vert
a_{i}b_{i}\right\rangle },  \label{Schmidt}
\end{equation}%
where $\sqrt{\mu _{i}}$ $(i=1,\ldots m)$ are the Schmidt coefficients, ${%
\left\vert a_{i}\right\rangle }$ and ${\left\vert b_{i}\right\rangle }$ are
orthonormal basis in $\mathcal{H}_{A}$ and $\mathcal{H}_{B}$, respectively.
It is evident that the reduced density matrices $\rho _{A}$ and $\rho _{B}$
have the same eigenvalues of $\mu _{i}$. It follows%
\begin{equation}
C^{2}({\left\vert \psi \right\rangle })=2\Big(1-\sum_{i}\mu _{i}^{2}\Big)%
=4\sum_{i<j}\mu _{i}\mu _{j},  \label{squareconcurrence}
\end{equation}%
which varies smoothly from $0$, for pure product states, to $2(m-1)/m$ for
maximally entangled pure states.

In order to derive a quantitative connection with the PPT criterion and the
realignment criterion, we recall some details of the two criteria. Peres
made firstly an important step forward for separability criterion in \cite%
{Peres96} by showing that $\rho ^{T_{A}}\geq 0 $ should be satisfied for a
separable state, where $\rho ^{T_{A}}$ stands for a partial transpose with
respect to the subsystem $A$. $\rho ^{T_{A}}\geq 0$ is further shown by
Horodecki \textit{et al.} \cite{HorodeckiPLA96} to be sufficient for $%
2\times 2$ and $2\times 3$ bipartite systems. $\Vert \rho ^{T_{A}}\Vert $ is
LU invariant as shown in Refs.\cite{Peres96,Vidal02} where $||\cdot ||$
stands for the trace norm defined by $\Vert G\Vert =Tr(GG^{\dagger })^{1/2}$%
. Thus it is sufficient to consider only the pure states with standard
Schmidt form given by Eq.~(\ref{Schmidt}). It is easy to see that $\rho
=\left\vert \psi \right\rangle \left\langle \psi \right\vert =\sum_{i,j}%
\sqrt{\mu _{i}\mu _{j}}{\left\vert a_{i}b_{i}\right\rangle }\left\langle
a_{j}b_{j}\right\vert $ and $\rho ^{T_{A}}=\sum_{i,j}\sqrt{\mu _{i}\mu _{j}}{%
\left\vert a_{j}^{\ast }b_{i}\right\rangle }\left\langle a_{i}^{\ast
}b_{j}\right\vert .$ Then we arrive at
\begin{eqnarray}
\Vert \rho ^{T_{A}}\Vert &=&\Vert \sum_{i,j}\sqrt{\mu _{i}\mu _{j}}{%
\left\vert a_{j}^{\ast }b_{i}\right\rangle }\left\langle b_{j}a_{i}^{\ast
}\right\vert \Vert  \nonumber \\
&=&\Vert \sum_{j}\sqrt{\mu _{j}}{\left\vert a_{j}^{\ast }\right\rangle }%
\left\langle b_{j}\right\vert \otimes \sum_{i}\sqrt{\mu _{i}}{\left\vert
b_{i}\right\rangle }\left\langle a_{i}^{\ast }\right\vert \Vert  \nonumber \\
&=&\Vert G\otimes G^{\dagger }\Vert =\Vert G\Vert ^{2}=(\sum_{i}\sqrt{\mu
_{i}})^{2}.  \label{rhoTA}
\end{eqnarray}%
where $G=\sum_{j}\sqrt{\mu _{j}}{\left\vert a_{j}^{\ast }\right\rangle }%
\left\langle b_{j}\right\vert $. In this derivation we have used the
unitarily invariant property of the trace norm when applying the elementary
column transformation: $\left\langle a_{i}^{\ast }b_{j}\right\vert
\rightarrow \left\langle b_{j}a_{i}^{\ast }\right\vert $ in the derivation
of the first formula. The last formula is obtained from the observation that
$GG^{\dagger }=\sum_{i,j}\sqrt{\mu _{i}\mu _{j}}{\left\vert a_{i}^{\ast
}\right\rangle }\left\langle b_{i}\right\vert \cdot {\left\vert
b_{j}\right\rangle }\left\langle a_{j}^{\ast }\right\vert =\sum_{i}\mu _{i}{%
\left\vert a_{i}^{\ast }\right\rangle }\left\langle a_{i}^{\ast }\right\vert
,$ the property of the trace norm $\Vert P\otimes Q\Vert =\Vert P\Vert \cdot
\Vert Q\Vert $ and the fact that $\Vert G\Vert $ is the sum of the square
root of eigenvalues $\mu _{i}$ of $GG^{\dagger }$.

Another complementary operational criterion for separability called the
\emph{realignment} criterion is very strong in detecting many of BES \cite%
{Rudolph02,ChenQIC03} and even genuinely tripartite entanglement \cite%
{Horodecki02}. Recently there has been considerable progress in the further
analysis, and in finding stronger variants and multipartite generalizations
for this criterion \cite{realignmentcriteria}. We recall that this criterion
states that a realigned version $\mathcal{R}(\rho )$ of $\rho $ should
satisfy $||\mathcal{R}(\rho )||\leq 1$ for any separable state $\rho $. $%
\mathcal{R}(\rho )$ is simply $\mathcal{R}(\rho )_{ij,kl}=\rho _{ik,jl}$
where $i$ and $j$ are the row and column indices for the subsystem $A$
respectively, while $k$ and $l$ are such indices for the subsystem $B$ \cite%
{Rudolph02,ChenQIC03,Horodecki02}. $||\mathcal{R}(\rho )||$ is also shown to
be LU invariant in \cite{ChenQIC03}. One has $\mathcal{R}(\rho )=\sum_{i,j}%
\sqrt{\mu _{i}\mu _{j}}{\left\vert a_{i}a_{j}^{\ast }\right\rangle }%
\left\langle b_{i}^{\ast }b_{j}\right\vert $ for the state Eq.~(\ref{Schmidt}%
), as follows easily from the definition. Similar to (\ref{rhoTA}) one has%
\begin{eqnarray}
\Vert \mathcal{R}(\rho )\Vert &=&\Vert \sum_{i}\sqrt{\mu _{i}}{\left\vert
a_{i}\right\rangle }\left\langle b_{i}^{\ast }\right\vert \otimes \sum_{j}%
\sqrt{\mu _{j}}{\left\vert a_{j}^{\ast }\right\rangle }\left\langle
b_{j}\right\vert \Vert  \nonumber \\
&=&\Vert G\otimes G^{\ast }\Vert =\Vert G\Vert ^{2}=(\sum_{i}\sqrt{\mu _{i}}%
)^{2}.  \label{realignment}
\end{eqnarray}%
where $G=\sum_{i}\sqrt{\mu _{i}}{\left\vert a_{i}\right\rangle }\left\langle
b_{i}^{\ast }\right\vert $. The last formula follows from the observation $%
GG^{\dagger }=\sum_{i,j}\sqrt{\mu _{i}\mu _{j}}{\left\vert
a_{i}\right\rangle }\left\langle b_{i}^{\ast }\right\vert \cdot {\left\vert
b_{j}^{\ast }\right\rangle }\left\langle a_{j}\right\vert =\sum_{i}\mu _{i}{%
\left\vert a_{i}\right\rangle }\left\langle a_{i}\right\vert $.

\vspace{0.05cm} We now derive the main result of this Letter.

\vspace{0.05cm} {Theorem:} \emph{For any $m\otimes n$ $(m\leq n)$ mixed
quantum state $\rho $, the concurrence $C(\rho )$ satisfies%
\begin{equation}
C(\rho )\geq \sqrt{\frac{2}{m(m-1)}}\Big(\max (\Vert \rho ^{T_{A}}\Vert
,\Vert \mathcal{R}(\rho )\Vert )-1\Big).  \label{maintheorem}
\end{equation}%
} \vspace{0.05cm}

\emph{Proof.---} To obtain the desired lower bound, let us assume that one
has already found an optimal decomposition $\sum_{i}p_{i}\rho ^{i}$ for $%
\rho $ to achieve the infimum of $C(\rho )$, where $\rho ^{i}$ are pure
state density matrices. Then $C(\rho )=\sum_{i}p_{i}C(\rho ^{i})$ by
definition. Noticing that $\Vert \rho ^{T_{A}}\Vert \leq \sum_{i}p_{i}\Vert
(\rho ^{i})^{T_{A}}\Vert $ and $\Vert \mathcal{R}(\rho )\Vert \leq
\sum_{i}p_{i}\Vert \mathcal{R}(\rho ^{i})\Vert $ due to the convex property
of the trace norm, one needs to show $C(\rho ^{i})\geq \sqrt{2/\big(m(m-1)%
\big)}(\Vert (\rho ^{i})^{T_{A}}\Vert -1)$ and $C(\rho ^{i})\geq \sqrt{2/%
\big(m(m-1)\big)}(\Vert \mathcal{R}(\rho ^{i})\Vert -1).$ For a pure state $%
\rho ^{i}$ one has $\Vert \mathcal{R}(\rho ^{i})\Vert =\Vert (\rho
^{i})^{T_{A}}\Vert =(\sum_{k}\sqrt{\mu _{k}})^{2}$ from Eqs.~(\ref{rhoTA})
and (\ref{realignment}), where $\sqrt{\mu _{k}}$ are the Schmidt
coefficients for the pure state $\rho ^{i}$. From the expression of Eq.~(\ref%
{squareconcurrence}) it remains to prove that%
\begin{eqnarray}
4\sum_{i<j}\mu _{i}\mu _{j} &\geq &\frac{2}{m(m-1)}((\sum_{k}\sqrt{\mu _{k}}%
)^{2}-1)^{2}  \nonumber \\
&=&\frac{8}{m(m-1)}(\sum_{i<j}\sqrt{\mu _{i}\mu _{j}})^{2},
\label{inequality1}
\end{eqnarray}%
where we have used $\sum_{i}\mu _{i}=1$.

The verification of the inequality Eq.~(\ref{inequality1}) is
straightforward: by summing over all of arithmetic mean inequalities $\mu
_{i}\mu _{j}+\mu _{k}\mu _{l}\geq 2\sqrt{\mu _{i}\mu _{j}\mu _{k}\mu _{l}}$
for $i<j$ and $k<l$, one gets%
\begin{eqnarray}
\sum_{i<j}\sum_{k<l}(\mu _{i}\mu _{j}+\mu _{k}\mu _{l}) &\geq
&2\sum_{i<j}\sum_{k<l}\sqrt{\mu _{i}\mu _{j}\mu _{k}\mu _{l}}  \nonumber \\
&=&2(\sum_{i<j}\sqrt{\mu _{i}\mu _{j}})^{2}.  \label{inequality2}
\end{eqnarray}%
It is seen that the number of appearance times is $m(m-1)$ for the term $\mu
_{i}\mu _{j}$ on the lhs of Eq.~(\ref{inequality2}). Therefore Eq.~(\ref%
{inequality1}) is confirmed and the conclusion Eq.~(\ref{maintheorem}) is
proved.

The most prominent feature of the Theorem is that it allows to obtain an
analytical lower bound for the concurrence without any numerical
optimization procedure. The bound has the same range as $C(\rho )$ and goes
from $0$ to $\sqrt{2(m-1)/m}$ for pure product states and maximally
entangled pure states, respectively. One can of course renormalizes the
maximum value $C(\rho )$ to be 1 with a change of the corresponding constant
factor.

We highlight some of the benefits of this new bound. Firstly, it serves to
detect and gives a lower bound of concurrence for \emph{all} entangled
states of two qubits and qubit-qutrit system. This is so because the PPT
criterion is necessary and sufficient for separability in the two cases \cite%
{HorodeckiPLA96}. Secondly, it generalizes to bipartite systems of arbitrary
dimension a relation given in \cite{Eisert99-Zyczkowski99-Verstraete01}
which is only valid for two qubits case, that the concurrence is lower
bounded by the negativity \cite{Vidal02} (defined to be $\Vert \rho
^{T_{A}}\Vert -1$). Thirdly, for any qubit-qudit system our bound can
contribute an analytical lower bound for EOF which is a convex function of
the concurrence, $E(\left\vert \psi \right\rangle )=H_2 \big((1+\sqrt{%
1-C^2(\left\vert \psi \right\rangle )})/2\big)$ where $H_2 (.)$ is the
binary entropy function \cite{Lozinski03}. In fact our bound can furnish a
lower bound of EOF $E(\rho)$ for arbitrary bipartite state $\rho$ \cite%
{Mintert04-MintertPhD}. Given any monotonously increasing, convex function $%
\mathcal{E}$ satisfying $\mathcal{E}(C({\left\vert \psi \right\rangle}))
\leq -\Sigma_r \mu_r \log_2 \mu_r$, one has $E(\rho)\geq \mathcal{E}(C(\rho))
$, with the rhs bounded from below by our bound Eq.~(\ref{maintheorem}).
Next we consider some examples to illustrate further the tightness and
significance of our bound.

\vspace{0.05cm} \emph{Example 1:} Isotropic states

Isotropic states \cite{Horodecki1999,Vollbrecht01} are a class of $U\otimes
U^{\ast }$ invariant mixed states in $d\times d$ systems
\begin{equation}
\rho _{F}={\frac{{1-F}}{d^{2}-1}}\left( I-|\Psi ^{+}\rangle \langle \Psi
^{+}|\right) +F|\Psi ^{+}\rangle \langle \Psi ^{+}|,  \label{isotropic}
\end{equation}%
where $|\Psi ^{+}\rangle \equiv \sqrt{1/d}\sum_{i=1}^{d}|ii\rangle $ and $%
F=\langle \Psi ^{+}|\rho _{F}|\Psi ^{+}\rangle $, satisfying $0\leq F\leq 1$%
, is the \emph{fidelity\/} of $\rho _{F}$ and $|\Psi ^{+}\rangle $. These
states were shown to be separable for $F\leq 1/d$~\cite{Horodecki1999}. It
is shown in \cite{Vidal02,Rudolph02} that $\Vert \rho _{F}^{T_{A}}\Vert
=\Vert \mathcal{R}(\rho _{F})\Vert =dF$ for $F>1/d$. The concurrence $C(\rho
)$ for this class of states is recently derived in \cite{Rungta03} to be $%
\sqrt{2d/(d-1)}(F-1/d)$ by an extremization procedure. An application of our
Theorem gives $C(\rho )\geq \sqrt{2/\big(d(d-1)\big)}(dF-1)=C(\rho )$. Thus
the bound gives surprisingly exact value of the concurrence for this sort of
states.

\noindent \emph{Remark:} One can see that the equality of Eq.~(\ref%
{inequality2}) holds when $\left\vert \psi \right\rangle$ are product states
or maximally entangled states (MES) (all $\mu_i$ are equal). Thus our bound
will be tight if an optimal decomposition for achieving concurrence only
involves product states and MES, and also attains the value of $\Vert \rho
^{T_{A}}\Vert$ or $\Vert \mathcal{R}(\rho )\Vert$. Roughly speaking, the
difference between our lower bound and the exact value of concurrence will
be small if there are few deviations from these two types of states in the
optimal ensemble decomposition. Exact estimation of this difference would be
an interesting subject for future study. In the case of isotropic states, it
is shown in \cite{Rungta03} that the optimal decomposition falls exactly
into this class and the concurrence is just our bound.

\vspace{0.05cm} \emph{Example 2:} $3\times 3$ BES constructed from
unextendible product bases (UPB)

In \cite{UPB}, Bennett \textit{et al.} introduced a $3\times 3$ BES from the
following bases:
\begin{eqnarray*}
{|\psi _{0}\rangle } &=&{\frac{1}{\sqrt{2}}}{|0\rangle }({|0\rangle }-{\
|1\rangle }),\ \ {|\psi _{1}\rangle }={\frac{1}{\sqrt{2}}}({|0\rangle }-{\
|1\rangle }){|2\rangle }, \\
{|\psi _{2}\rangle } &=&{\frac{1}{\sqrt{2}}|2\rangle }({|1\rangle }-{\
|2\rangle }),\text{ \ }{|\psi _{3}\rangle }={\frac{1}{\sqrt{2}}}({|1\rangle }%
-{|2\rangle }){|0\rangle }, \\
{|\psi _{4}\rangle } &=&{\frac{1}{3}}({|0\rangle }+{|1\rangle }+{|2\rangle )}%
({|0\rangle }+{|1\rangle }+{|2\rangle ),}
\end{eqnarray*}%
from which the density matrix could be expressed as
\begin{equation}
\rho =\frac{1}{4}(Id-\sum_{i=0}^{4}{|\psi _{i}\rangle \langle \psi _{i}|}).
\label{upb1}
\end{equation}%
A simple calculation gives $\Vert \rho ^{T_{A}}\Vert =1$ and $\Vert \mathcal{%
R}(\rho )\Vert =1.087$ \cite{ChenQIC03}, therefore $C(\rho )\geq 0.05$
according to the Theorem. This shows that the state is entangled.

When BES are constructed from the UPB \cite{UPB} given by $|\psi _{j}\rangle
=|\vec{v}_{j}\rangle \otimes |\vec{v}_{2j\bmod5}\rangle ,\;\;(j=0,\ldots ,4)$
with $\vec{v}_{j}=N(\cos (2\pi j /5),\sin (2\pi j /5),h)$, with $j=0,\ldots
,4$, $h=\sqrt{1+\sqrt{5}}/2$ and $N=2/\sqrt{5+\sqrt{5}}$, then the PPT state
of Eq.~(\ref{upb1}) gives $\Vert \rho ^{T_{A}}\Vert =1$ and $\Vert \mathcal{R%
}(\rho )\Vert =1.098 $ \cite{ChenQIC03}, therefore $C(\rho )\geq 0.056$
according to the Theorem, which identifies this BES.

It is conjectured by Audenaert \textit{et al.} that the optimization method
for concurrence is a necessary and sufficient for separability when one
considers all possible complex linear combination of the concurrence-vectors
\cite{Audenaert01}. Our numerical verification suggests a disproval for
this conjecture, because of the failure to identify entanglement by applying their
optimization method for the above two UPB states. Thus the direct estimation
method for concurrence in \cite{Audenaert01} may not be able to detect \emph{all}
entangled states through numerical optimizations. Here our Theorem
complements other existing approaches to make a quite good estimate of
entanglement for BES.

\vspace{0.05cm} \emph{Example 3:} Horodecki's $3\times 3$ entangled state

A mixed two qutrits is introduced in \cite{HorodeckiPRL99}:%
\begin{equation}
\sigma _{\alpha }=\frac{2}{7}|\Psi ^{+}\rangle \langle \Psi ^{+}|+\frac{%
\alpha }{7}\sigma _{+}+\frac{5-\alpha }{7}\sigma _{-},  \label{3hstate}
\end{equation}%
where
\begin{eqnarray}
\sigma _{+} &=&{\frac{1}{3}}(|0\rangle |1\rangle \langle 0|\langle
1|+|1\rangle |2\rangle \langle 1|\langle 2|+|2\rangle |0\rangle \langle
2|\langle 0|),  \nonumber \\
\sigma _{-} &=&{\frac{1}{3}}(|1\rangle |0\rangle \langle 1|\langle
0|+|2\rangle |1\rangle \langle 2|\langle 1|+|0\rangle |2\rangle \langle
0|\langle 2|),  \nonumber \\
\left\vert \Psi ^{+}\right\rangle &=&\frac{1}{\sqrt{3}}\left( |0\rangle
|0\rangle +|1\rangle |1\rangle +|2\rangle |2\rangle \right) .
\label{3hbasis}
\end{eqnarray}%
In \cite{HorodeckiPRL99} Horodecki \textit{et al.} demonstrate that the
states Eq. (\ref{3hstate}) admit a simple characterization with respect to
the parameter $2\leq \alpha \leq 5$: separable for $2\leq \alpha \leq 3$;
bound entangled for $3<\alpha \leq 4$; free entangled for $4<\alpha \leq 5$.
It is computed by using the realignment criterion in \cite{Rudolph02} that $%
\Vert \mathcal{R}(\sigma _{\alpha })\Vert =(19+2 \sqrt{3\alpha ^{2}-15\alpha
+19})/21$ and one can recognize all the entangled states for $3<\alpha \leq
5.$ One can obtain further that $\Vert \sigma _{\alpha }^{T_{A}}\Vert =1$
for $2\leq \alpha \leq 4$ and $\Vert \sigma _{\alpha }^{T_{A}}\Vert =(2+%
\sqrt{4\alpha ^{2}-20\alpha +41})/7$ for $4<\alpha \leq 5.$ Therefore one
has $C(\sigma _{\alpha })\geq 1/\sqrt{3}(\Vert \mathcal{R}(\sigma _{\alpha
})\Vert -1)=2 \sqrt{3}(\sqrt{3\alpha ^{2}-15\alpha +19}-1)/63$ due to the
observation that $\Vert \mathcal{R}(\sigma _{\alpha })\Vert $ is always
greater than $\Vert \sigma _{\alpha }^{T_{A}}\Vert $ in the entangled region
$3<\alpha \leq 5$.

However, the concurrence optimization procedure proposed in \cite%
{Mintert04-MintertPhD} can only identify the entangled states for $%
3.52\lesssim \alpha \leq 5$ \cite{Mintert04-MintertPhD}. This
suggests that Mintert \textit{et al.}'s methods may not be
necessary and sufficient for
detecting entanglement. A rough comparison with the result of \cite%
{Mintert04-MintertPhD} shows that our lower bound is much better
than their optimized bound in the entangled region of $3<\alpha
\lesssim 4.75 $, though a little bit weaker than theirs in the
region $4.75\lesssim \alpha \leq 5$.

We remark that, like any other known approaches, there are also
some drawbacks for our estimation. Our lower bound cannot detect all the
entangled states due to limitation of the PPT criterion and the
realignment criterion. For example, it can neither recognize the
$2\times 4$ Horodecki BES \cite{HorodeckiPLA97}, which instead can
be detected by the methods of \cite{Mintert04-MintertPhD}, nor
give the exact value of concurrence for 2 qubits known from
\cite{Wootters98,Mintert04-MintertPhD}.

In summary, we have provided an entirely analytical formula for
lower bound of concurrence, by making a novel connection with the
known strong separability criteria. The bound leads to actual
values of concurrence for some special class of quantum states.
One only needs to calculate the trace norm of certain matrices,
which avoids complicated optimization procedure over a large number
of free parameters in numerical approaches. The formula
also permits to furnish lower bounds of EOF for arbitrary
bipartite quantum state. This complements the nice result of
Wootters for 2 qubits, as well as a number of existing
optimization methods for concurrence. Profiting from the strong
realignment criterion, our bound can give easy entanglement
evaluation for many BES, which fail to be recognized by the
regular optimization methods. This shows that our method can serve
as a powerful tool for investigating both static and dynamical
entanglement properties in realistic quantum computing devices. As
applications the method could be used in indicating a possible
quantum phase transition for condensed matter system, and in
analyzing finite size or scaling behavior of entanglement in
various interacting quantum many-body systems.

K.C. gratefully acknowledges support from the Alexander von Humboldt
Foundation. This work has been supported the Deutsche Forschungsgemeinschaft
SFB611 and German(DFG)-Chinese(NSFC) Exchange Programme 446CHV113/231. We
thank Zhi-Xi Wang for valuable discussions.

\end{document}